\documentclass[a4paper,11pt]{article}
\usepackage{pos}
\usepackage{float}
\title{
The upper right corner of the Columbia plot with staggered fermions
 }
\author*[a]{Ruben Kara}
\author[a]{Szabolcs Bors\'anyi}
\author[a,b,c,d]{Zoltán Fodor}

\author[a]{Jana N. Guenther}
\author[a,b]{Paolo Parotto}
\author[d]{Attila P\'asztor}
\author[e]{Dénes Sexty}

\affiliation[a]{University of Wuppertal, Department of Physics, Wuppertal D-42119, Germany}

\affiliation[b]{Pennsylvania State University, Department of Physics, State College, PA 16801, USA}

\affiliation[c]{Jülich Supercomputing Centre, Forschungszentrum Jülich, Jülich D-542425, Germany}
\affiliation[d]{E{\"o}tv{\"o}s University, Budapest 1117, Hungary}
\affiliation[e]{University of Graz, Department of Physics, Graz A-8010, Austria}

\emailAdd{rkara@uni-wuppertal.de}

\abstract{QCD with heavy dynamical quarks exhibits a first order thermal transition which is driven by the spontaneous breaking of the global $\mathcal{Z}_3$ center symmetry. Decreasing the quark masses weakens the transition until the corresponding latent heat vanishes at the critical mass.
We explore the heavy mass region with three flavors of staggered quarks and analyze the Polyakov loop and its moments in a finite volume scaling study. We calculate the heavy critical mass in the three flavor theory in the infinite volume limit for $N_t=8$.}

\FullConference{%
 The 38th International Symposium on Lattice Field Theory, LATTICE2021
  26th-30th July, 2021
  Zoom/Gather@Massachusetts Institute of Technology
}

\begin{document}
\maketitle

\section{Introduction}
\noindent
It is well known that for physical quark masses and vanishing chemical potential $\mu=0$ the QCD transition is an analytic crossover \cite{Aoki:2006we}. In contrast to that, for infinite heavy quark masses, QCD exhibits a first order thermal transition due to the  spontaneous breaking of the global $\mathcal{Z}_3$ center symmetry. The corresponding latent heat was calculated recently in the continuum limit \cite{Shirogane:2016zbf,Shirogane:2020muc}.\\
Several approaches to study criticality in QCD in the heavy mass region, such as the derivative method, flow time expansion, reweighting from quenched QCD, the hopping parameter expansion or effective actions including the Polyakov loop term have been used and combined \cite{Shirogane:2016zbf,Shirogane:2020muc,Hiroshi:2013,Reweight,Kiyohara:2021smr}. Important results are the critical hopping parameter $\lambda_c$ and the latent heat or the energy gap (quenched QCD), clearly indicating a first order phase transition.\\
For vanishing quark masses, the so-called chiral limit, QCD possesses the global $SU(3)$-chiral symmetry. This symmetry is spontaneously broken at low temperatures and restored at higher temperatures, thus it shows the opposite behavior compared to the center symmetry. The order of the transition cannot be determined via direct simulations in the chiral limit, but the $\epsilon-$expansion in the linear sigma model indicates a first order phase transition in three dimensions for $N_f \geqslant 3$ \cite{Wilczek1984}. On the other hand, the latest simulations at finite lattice spacings are compatible with the scenario without a first order corner in the chiral region \cite{Cuteri:2021ikv, Dini:2021hug}.
\begin{center}
\includegraphics[width=0.51\textwidth]{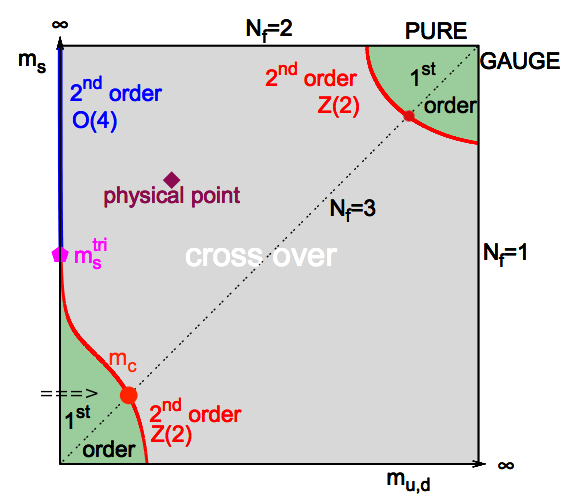}
\captionof{figure}{Columbia plot for the $N_f=2+1$ flavor theroy. We focus on the search of the critical endpoint in the heavy mass region on the diagonal. Figure from \cite{deForcrand:2017cgb}.}
\label{fig:columbia}
\end{center}
For finite quark masses, both center and chiral symmetry are explicitly broken and Monte Carlo simulations show that for intermediate masses the transition is analytic. The two first order regions must be separated from the crossover region by second order lines. The universality class of these critical endpoints is that of the 3d Ising model. For vanishing chemical potential the phase diagram is summarized in the so-called Columbia plot, shown in fig. \ref{fig:columbia}, where the quark masses play the role of the external parameters.\\
In this proceedings we focus on the determination of the critical endpoint in form of a critical mass $m_c$ in the upper-right corner of the Columbia plot for $N_f=3$, i.e. along the diagonal. For this purpose we perform simulations in the three flavor theory for a large set of masses and couplings.

\section{Analysis}
\noindent
A natural choice for the observables to probe the center symmetry breaking are the Polyakov loop $P$ and its susceptibility $\chi$. $P$ transforms non-trivially under $Z_3$ and both quantities are defined as
\begin{equation}
P=\frac{1}{N_s^3} \sum_{\vec{x}} P_{\vec{x}} = \frac{1}{N_s^3} \sum_{\vec{x}} \mathrm{tr} \left[ \prod_\tau U_4(\vec{x},\tau)   \right] \qquad \chi=N_s^3 \left( \langle |P|^2 \rangle   - \langle |P| \rangle^2 \right),
\end{equation}
where $N_s$ stands for the spatial extension of the lattice, while $\vec{x}$ and $\tau$ indicate the spatial and the temporal position respectively.\\
Since we simulate systems near a critical endpoint, we have to deal with critical slowing down and a diverging correlation length. In order to reduce the auto-correlation time, we employ parallel tempering in $\beta$ \cite{Marinari1992}, thus performing multiple simulations at different couplings.
%Since we simulate highly critical systems we face critical phenomena such as a diverging correlation length. Thus we use parallel tempering in $\beta$ to reduce the auto-correlation time \cite{Marinari1992}.
\begin{center}
\includegraphics[width=0.65\textwidth]{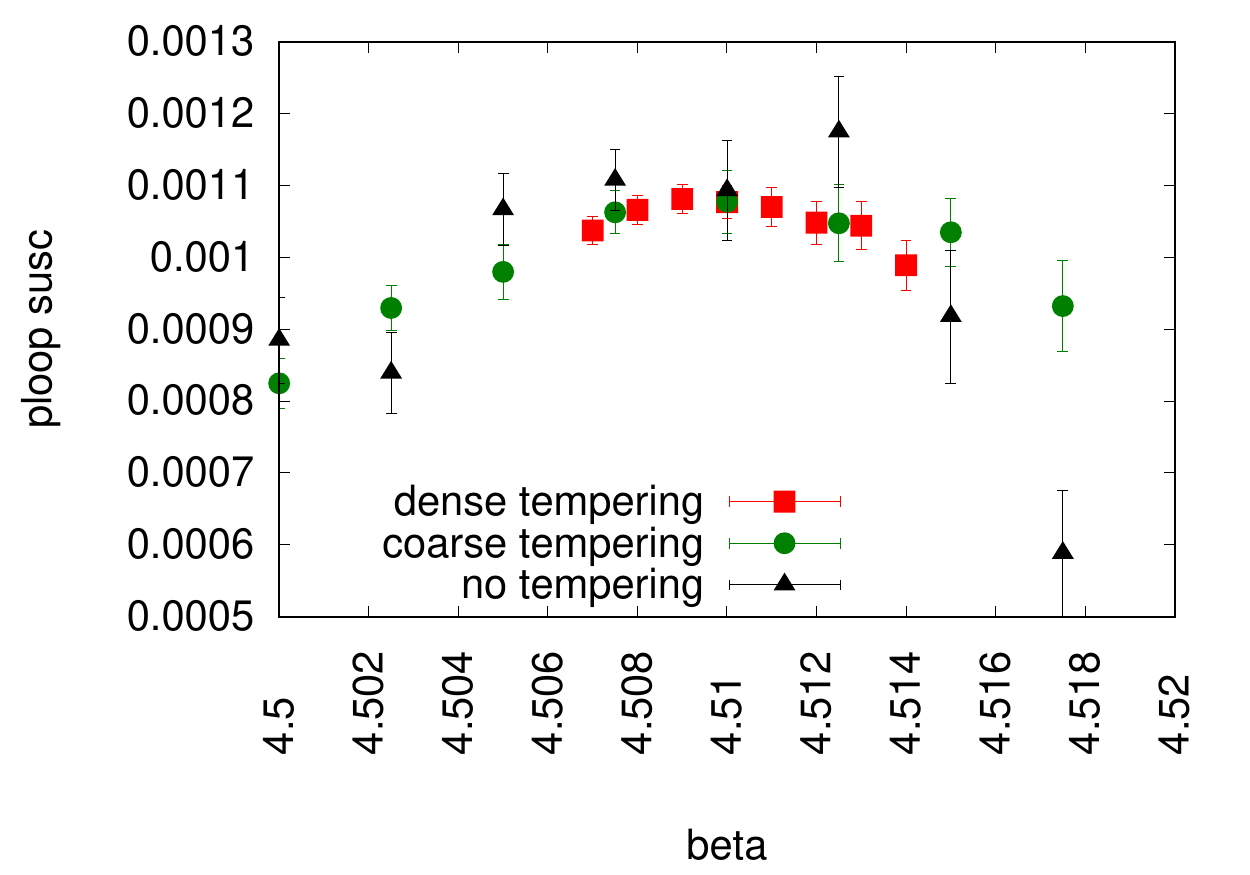}
\captionof{figure}{Polyakov loop susceptibility ($\chi/N_s^3$) as a function of the coupling $\beta$ in quenched QCD. The simulations were performed on a $32^3 \times 8$ lattice. Each data set was simulated on 8 KNL cards for 24 hours. The black triangles are the results of brute-force overrelaxation/heatbath simulations. The green circles and the red squares are both results from the tempered simulations with two different spacings between the $\beta$ values.}
\label{fig:tempering}
\end{center}
Parallel tempering takes advantage of the fact that these simulations are distinct Markov processes whose equilibrium distributions overlap. Swapping configurations between pairs of sub-ensembles according to a Metropolis accept/reject step allows us to reduce the auto-correlation time within this pair of simulations. The probability of these transitions strongly depends on the difference between the parameter set of the simulations, i.e. swapping configurations is more likely for neighboring $\beta$ ensembles. The price to pay is a resulting correlation between the simulations.
 %So the configurations can be swapped and the auto-correlation time be reduced. The price to pay is a correlation between the ensembles at different $\beta$. %This will lead to a reduced auto-correlation time for the price of a correlation between the ensembles at different $\beta$.
Further details can be found in \cite{joo1998}. In fig. \ref{fig:tempering} the advantages of this method are clearly visible, since the errorbars are significantly smaller in the case of tempering compared to the standard algorithm.\\
The Polyakov loop serves as a true order parameter for the thermal center symmetry breaking in the case of quenched QCD. A vanishing value of $\langle |P| \rangle$ in the infinite volume limit indicates confinement and $\langle |P| \rangle \neq 0$ deconfinement at higher $T$, since it is linked to the static quark potential \cite{Benjamin}. For dynamical simulations the explicit breaking of $\mathcal{Z}_3$ always leads to a non-vanishing value of $\langle |P| \rangle$. Nevertheless, the Polyakov loop is still a steeply rising function in the transition region and its inflection point could be associated with the transition temperature. Instead, we focus on the peak of its susceptibility. In fig. \ref{fig:polya_beta} both quantities are shown as functions of the coupling.
\begin{center}
\includegraphics[width=0.51\textwidth]{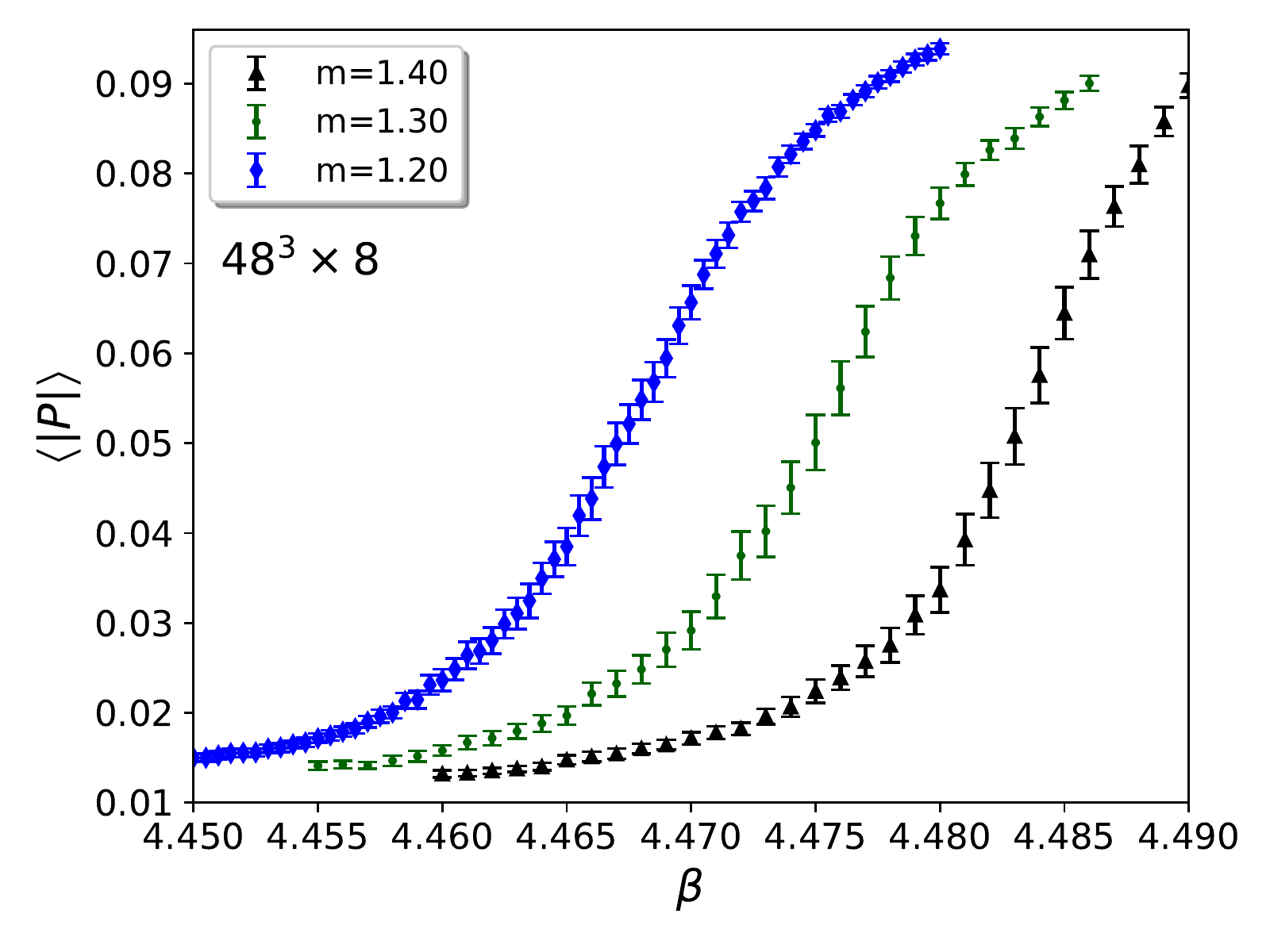}\includegraphics[width=0.51\textwidth]{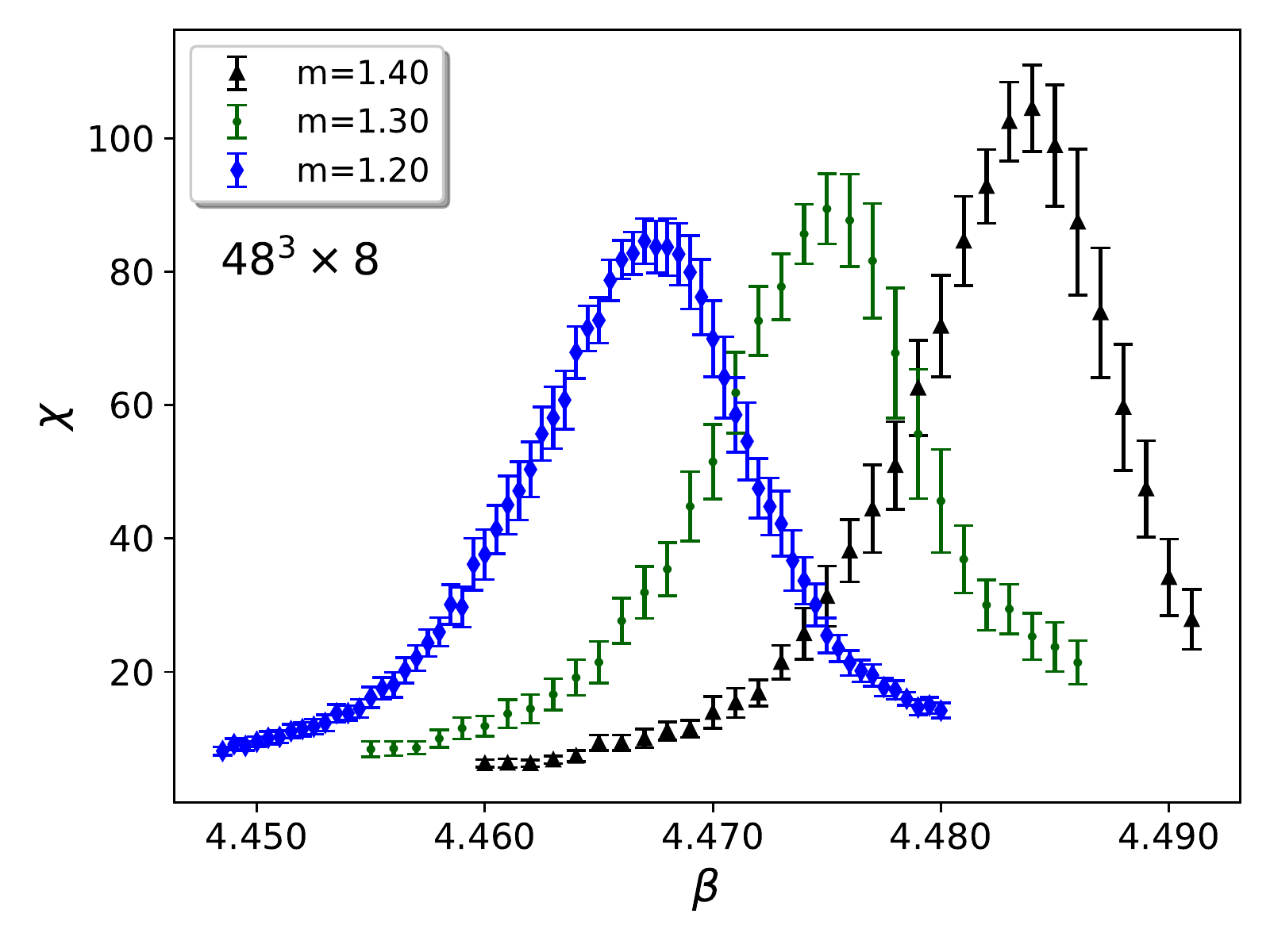}
\captionof{figure}{Polyakov loop (left) and its susceptibility (right) as a function of the coupling $\beta$ at several masses in the vicinity of the critical mass for a given volume.}
\label{fig:polya_beta}
\end{center}
\noindent
Analyzing the peak of the susceptibility is a suitable method to determine the type of phase transition. A diverging peak in the infinite volume limit is a clear sign of a real transition. The main task is then how to determine the peak as precisely as possible. First we express the susceptibility as a function of the Polyakov loop, as shown in fig. \ref{fig:chi_polya}. This has the advantage that the form of $\chi(\langle |P| \rangle)$ is simpler compared to $\chi(\beta)$ (see fig. \ref{fig:polya_beta}), whereby the latter is compatible with a larger set of possible fitting functions.
\begin{center}
\includegraphics[width=0.51\textwidth]{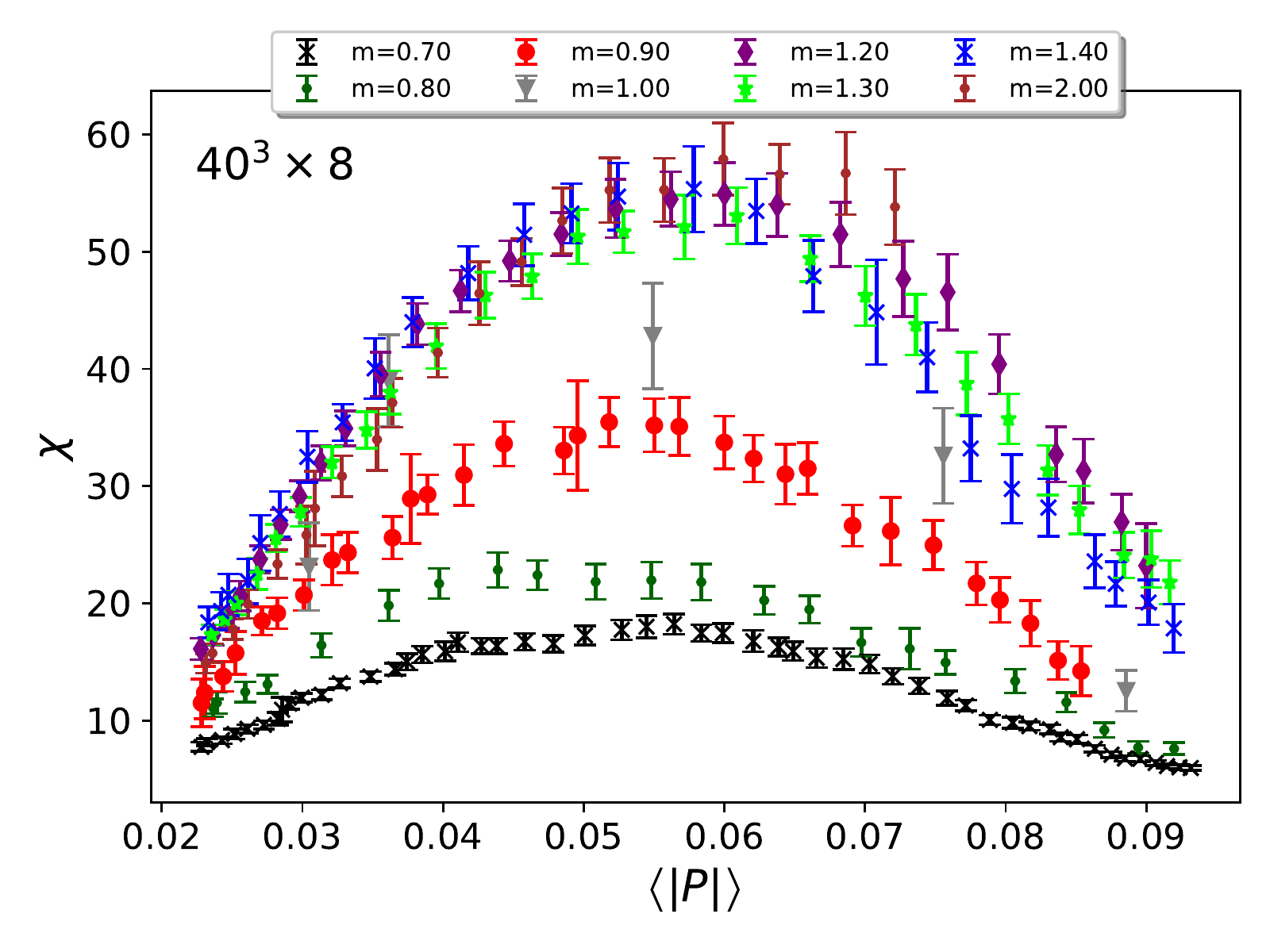}\includegraphics[width=0.51\textwidth]{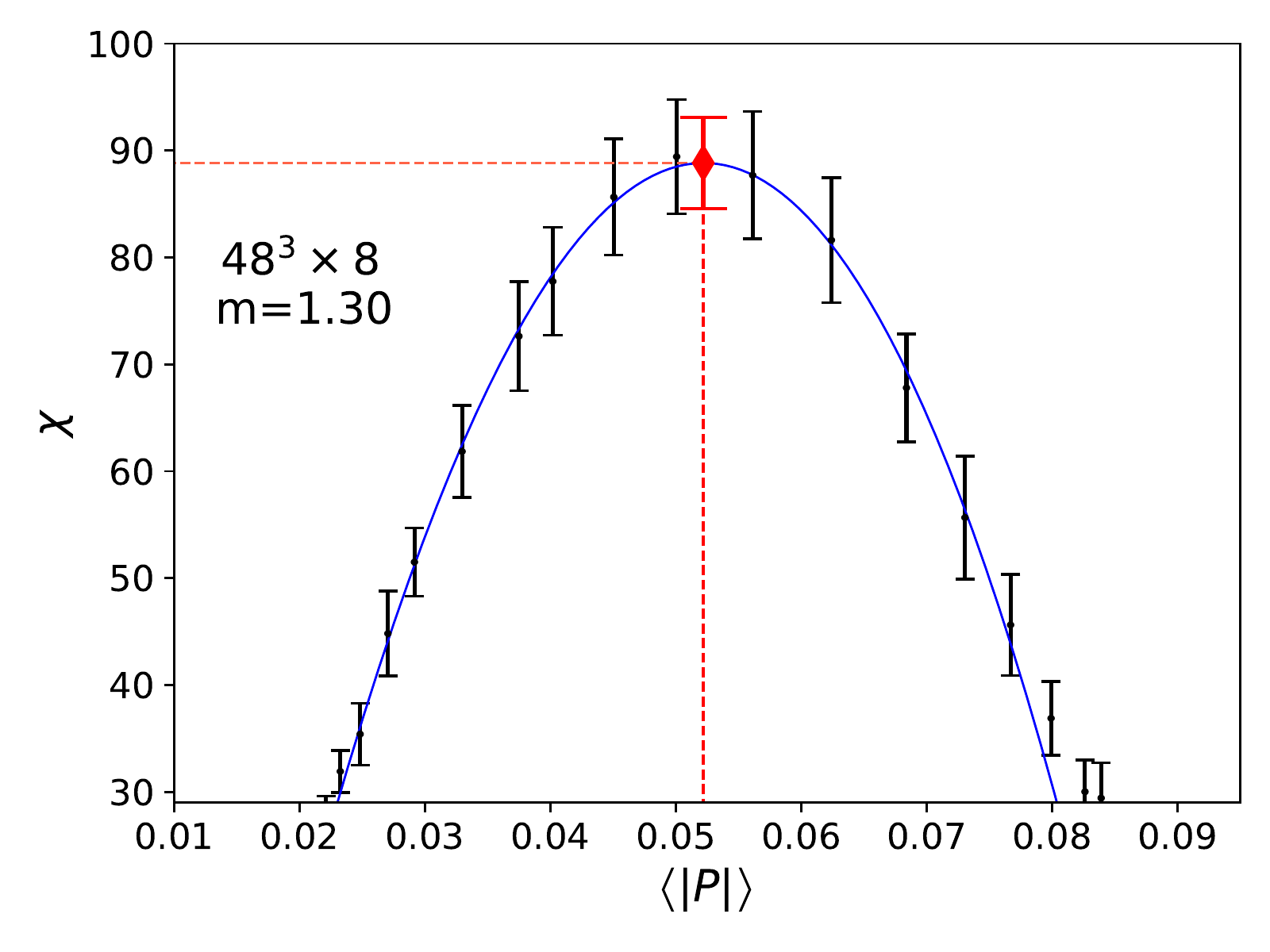}
\captionof{figure}{Left: Susceptibilities as function of the Polyakov loop for sevaral masses on a $40^3\times 8$ lattice. Right: Illustration of the peak determination via a low order polynomial fit.}
\label{fig:chi_polya}
\end{center}
\noindent
For this purpose $\langle |P| \rangle(\beta)=\mathrm{const.}$ is solved and $\beta$ substituted into $\chi(\beta)$. The statistical error on $\langle |P| \rangle$ is converted into an additional error on $\chi$. We used a similar strategy in \cite{chira} in which we expressed the chiral susceptibility as a function of the chiral condensate.\\
In the left panel of fig. \ref{fig:chi_polya} we can observe that the peak is only weakly mass dependent. Its height grows by increasing the quark masses, but this tendency gets weaker for the higher masses.\\

\section{Volume scaling of $\boldsymbol{\chi_\mathrm{max}}$}
\noindent
In the case of a first order phase transition the maximum of the susceptibility $\chi_\mathrm{max}$ diverges linearly with the physical volume. For a second order phase transition this behavior is accompanied by a critical exponent. In contrast to that, an analytic crossover shows a finite value of $\chi_\mathrm{max}$ in the infinite volume limit. In fig. \ref{fig:volscaling} we present the inverse physical volume $1/(LT_c)^3$ scaling of the inverse peak for some values of $m$.
\begin{center}
\includegraphics[width=0.65\textwidth]{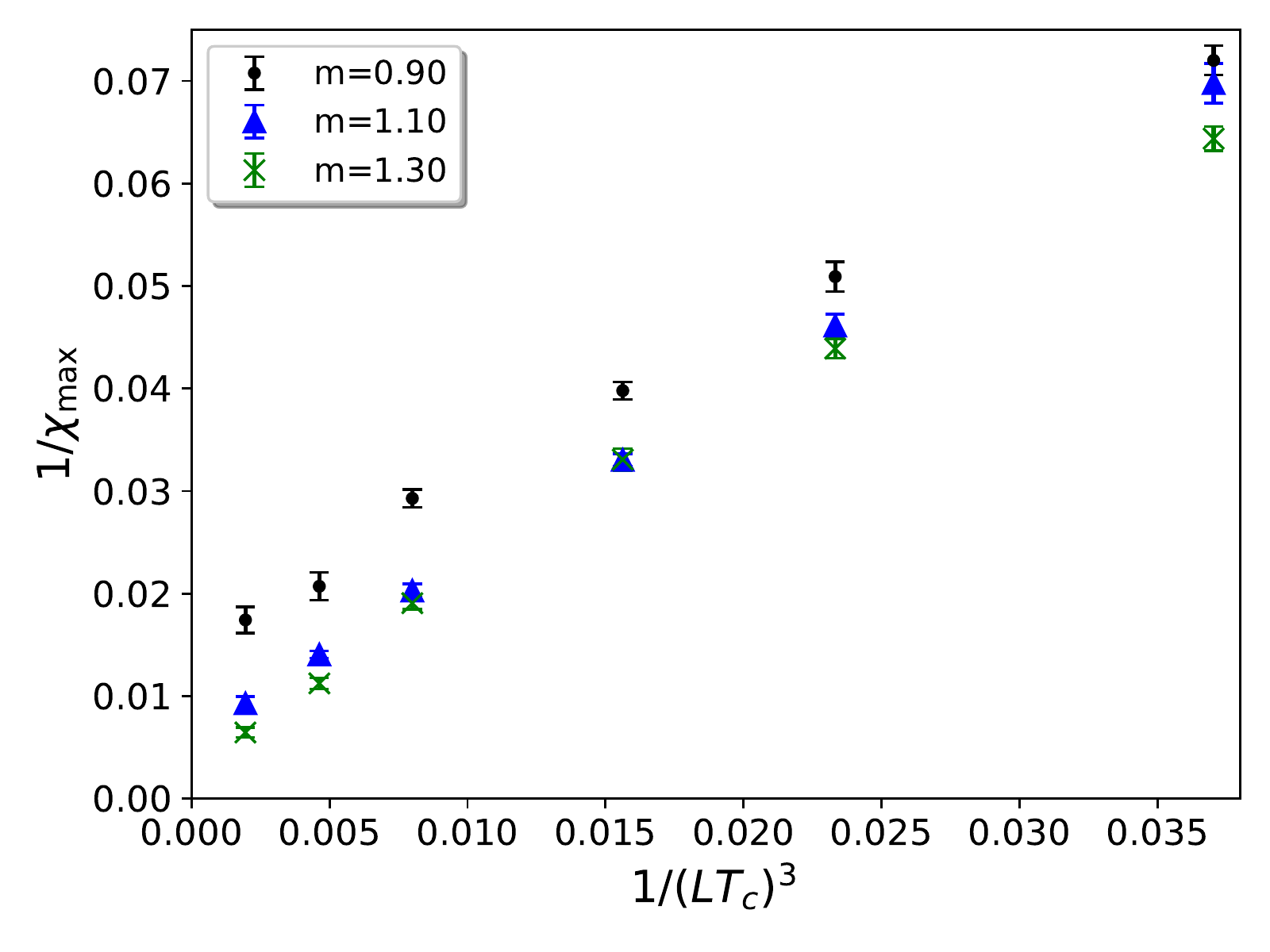}
\captionof{figure}{Inverse susceptibility peak as a function of the inverse physical volume for $m=0.90,1.10,1.30$.}
\label{fig:volscaling}
\end{center}
\noindent
For $m=0.90$ the inverse susceptibility peak takes a finite and non-vanishing value in the infinite volume limit, which is a clear sign that the system is in the crossover region. By increasing the mass the system seems to get more and more critical, since $\chi_\mathrm{max}^{-1}$ tends to vanish for $(LT_c)^3 \rightarrow \infty$, but not strictly linearly.\\
To investigate the systematics of the analysis, we initially extract the infinite volume limit via a linear fit or via a linear fit with an "effective exponent", defined as
\begin{equation}
\chi^{-1} \left(LT_c\right)= a+b\cdot \left( \frac{1}{LT_c^3} \right) ^{c},
\label{eq:inf_vol}
\end{equation}
\noindent
where $a,b,c$ are fitting parameters.

\section{Determination of the critical mass}
\noindent
In the case of a second order phase transition, $\chi_\mathrm{max}^{-1}$ follows a power law near the transition. Thus, we expect the inverse susceptibility peak to vanish in the infinite volume limit according to a power law, in which the quark mass represents the symmetry breaking field. The specific power is given by the Ising critical exponent $\gamma$. We therefore calculate the critical mass $m_c$ through a fit via
\begin{equation}
\chi_\mathrm{max}^{-1}\left(LT_c \rightarrow \infty \right)= A\cdot (m_c -m)^\gamma ,
\label{eq:crit_mass}
\end{equation}
\noindent
with $A$ and $m_c$ as fitting parameters and $\gamma=1.2373$ \cite{Ising}.

\begin{center}
\includegraphics[width=0.51\textwidth]{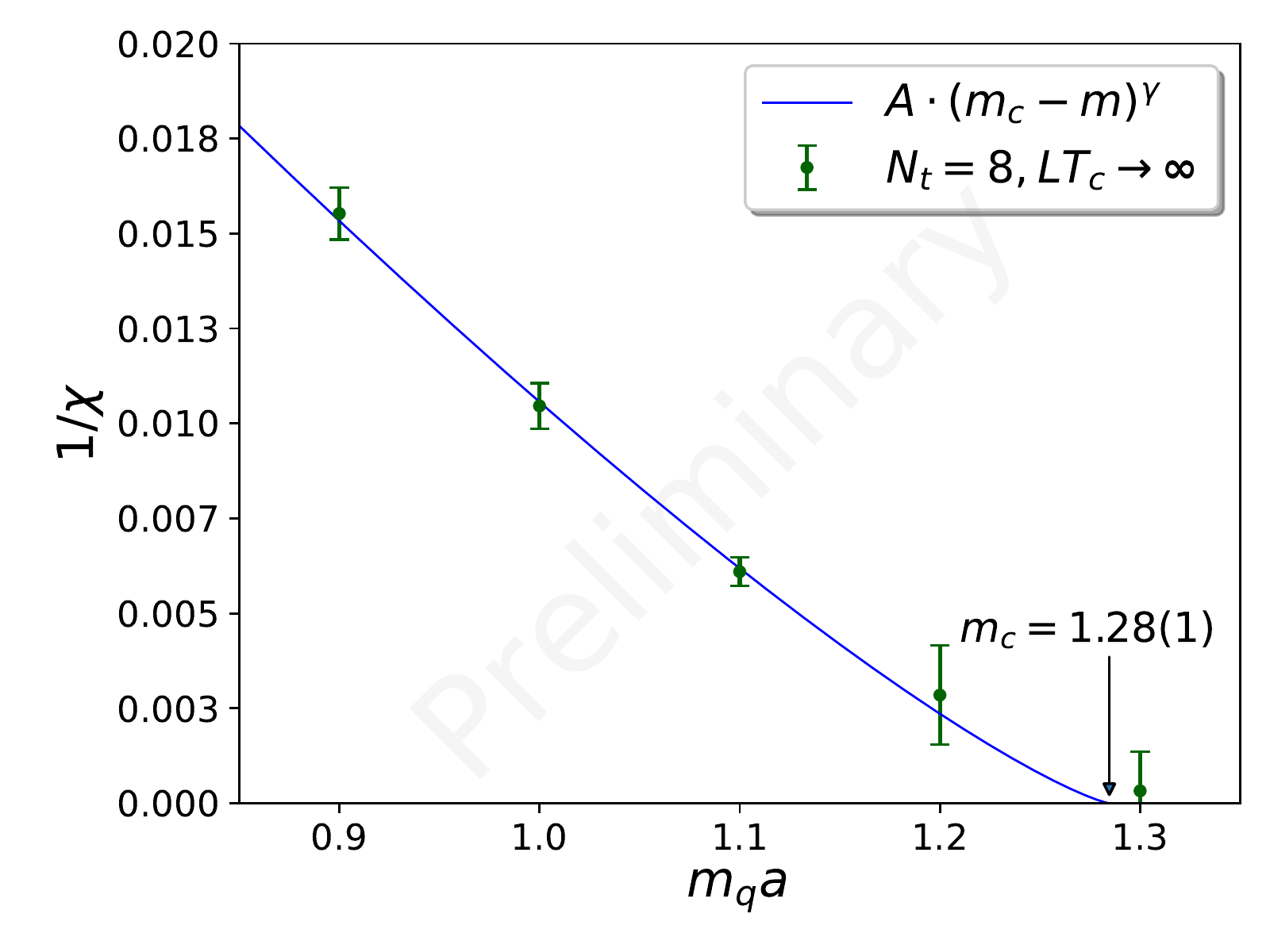}\includegraphics[width=0.51\textwidth]{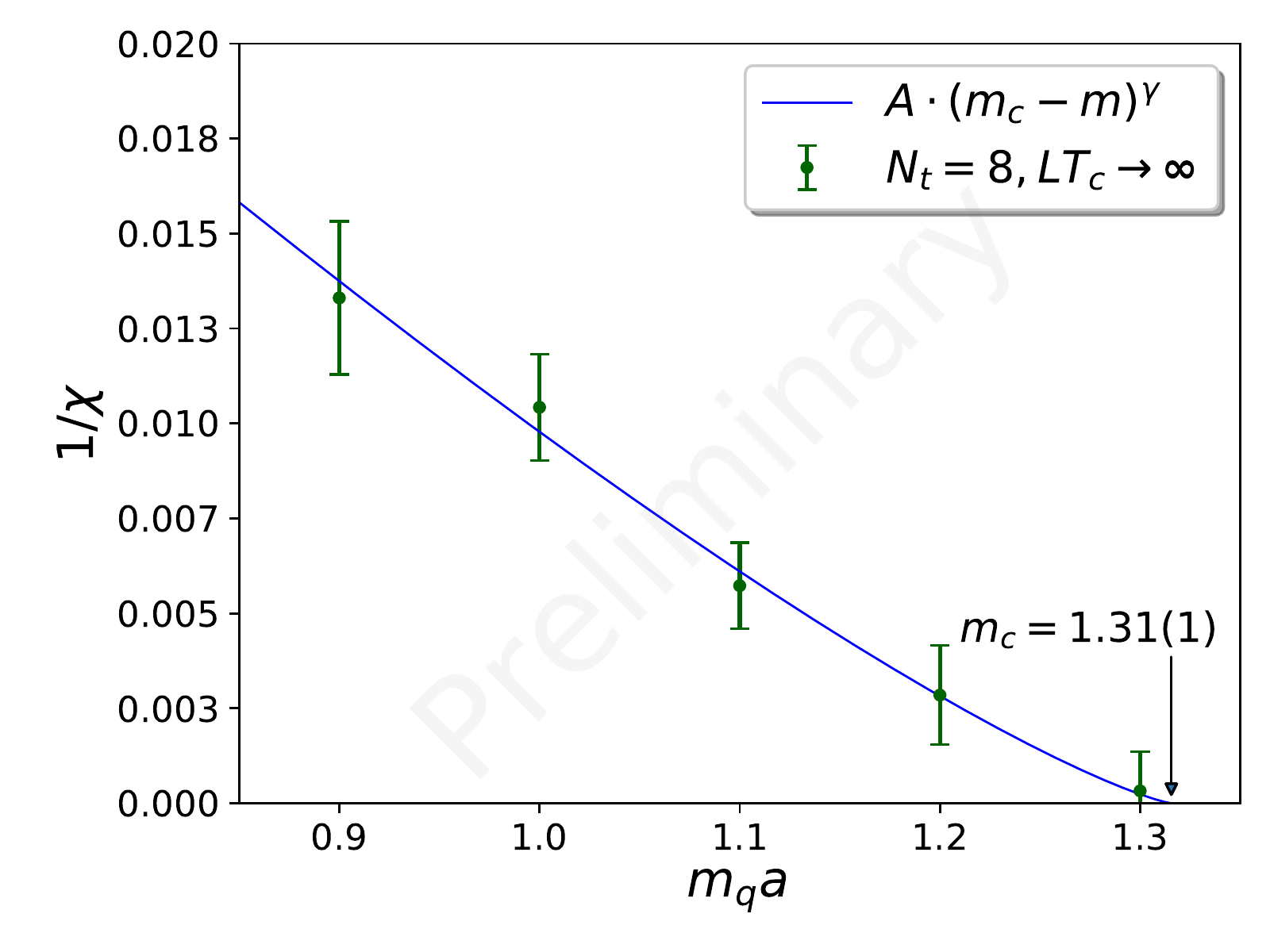}
\captionof{figure}{Inverse susceptibility peak in the infinite volume limit as a function of the quark mass.\\ Left: Infinite volume limit calculated via linear fits for $m=0.9 , 1.0, 1.1$ and $m=1.2, 1.3$ via linear fits with effective exponent.\\
Right: Infinite volume limit calculated via linear fits with effective exponent.}
\label{fig:crit_mass}
\end{center}
\noindent
The results of the fits to $\chi_\mathrm{max}^{-1}$ are shown in fig. \ref{fig:crit_mass}. For each mass the infinite volume limit was taken via eq.~(\ref{eq:inf_vol}). In the left panel, however, we fix the effective exponent $c=1$ for the smallest three masses $m=0.9 , 1.0, 1.1$. From these two analyses we obtain two critical masses $am_c=1.28(1)$ (left panel) and $am_c=1.31(1)$ (right panel).\\
The critical masses in fig. \ref{fig:crit_mass} are given in lattice units. To relate them to physical quantities, we calculate the pseudoscalar mass $m_\mathrm{PS}$ and the scale parameter $w_0$ \cite{Borsanyi:2012zs}. Therefore we construct dimensionless quantities with the critical temperature $T_c$ and obtain $m_\mathrm{PS}/T_c=19.1(1)$ and $w_0 T_c = 0.2507(2)$ for $m_c=1.28(1)$.

\section*{Acknowledgements}
\noindent
A.P. is supported by the
J\'anos Bolyai Research Scholarship of the Hungarian Academy of Sciences and by
the UNKP-21-5 New National Excellence Program of the Ministry of Innovation and
Technology.
The authors gratefully acknowledge the Gauss Centre for
Supercomputing e.V. (www.gauss-centre.eu) for funding this project by providing
computing time on the GCS Supercomputer JUWELS and JURECA/Booster at J\"ulich
Supercomputing Centre (JSC). Parts of the computations were performed
on the QPACE3 system, funded by the DFG.

%In this proceedings we set the focus on the determination of the critical mass along the diagonal of the Columbia plot in the heavy mass region. To find out the type of phase transition, the precise calculation of the susceptibility peak is necessary. Therefore we expressed the susceptibility as a function of the Polyakov loop to get a simpler form compared to $\chi(\beta)$. Then the peak can be extracted via a low order polynomial fit. The infinite volume limit is then calculated by a linear fit or a linear fit with an effective exponent (eq.~(\ref{eq:inf_vol})). In the case of a second order phase transition, $\chi_\mathrm{max}^{-1}$ diverges according to a power law. Using the Ising critical exponent $\gamma$, the critical mass is calculated to $m_c=1.28(1)$ or $m_c=1.31(1)$. Constructing dimensionless quantities leads to $m_\mathrm{PS}/T_c=19.1(1)$ and $\omega_0 T_c = 0.2507(2)$ for $m_c=1.28(1)$.

%\begin{center}
%\includegraphics[width=0.51\textwidth]{plots/mps_crit.pdf}\includegraphics[width=0.51\textwidth]{plots/w0_crit.pdf}
%\captionof{figure}{Calculation of the dimensionless quantities $\omega_0 T_c$ and $m_\mathrm{PS}/T_c$ for $m_c=1.28(1)$.}
%\label{fig:omega}
%\end{center}
%\noindent
%The pseudoscalar mass is calculated to $m_\mathrm{PS}/T_c=19.1(1)$ and $\omega_0 T_c = 0.2507(2)$.

\end{document}